\documentclass[
twocolumn,
superscriptaddress,
preprintnumbers,
prl,
]{revtex4-2}

\usepackage{amsmath,amssymb}
\usepackage{graphicx}
\usepackage{dcolumn}
\usepackage{bm}
\usepackage{float}
\usepackage[dvipsnames]{xcolor}
\usepackage{hyperref}
\hypersetup{colorlinks=true,allcolors=MidnightBlue}
\usepackage{braket}  
\usepackage{xcolor}

\begin{document}

\title{The tilted-plane structure of the energy of 
  {finite} quantum systems}

\author{Andrew C. Burgess}
\affiliation{%
School of Physics, Trinity College Dublin, The University of Dublin, Ireland} 

\author{Edward Linscott}%
\affiliation{Laboratory for Materials Simulations, Paul Scherrer Institut, 5232
Villigen PSI, Switzerland}
\affiliation{National Centre for Computational Design and Discovery of Novel
Materials (MARVEL), Paul Scherrer Institut, 5232 Villigen PSI, Switzerland}

 \author{David D. O'Regan}%
 \email{david.o.regan@tcd.ie}
\affiliation{%
School of Physics, Trinity College Dublin, The University of Dublin, Ireland}

\date{\today}

\begin{abstract}
The piecewise linearity condition
on the total energy with respect to   {the} total magnetization  of   {finite}  quantum
systems is derived,  using the infinite-separation-limit technique. 
This generalizes the well-known
constancy condition, related
to static correlation error, in 
approximate density functional theory
(DFT).
The magnetic analog of the DFT Koopmans' theorem is also derived.
Moving to fractional electron count, 
the tilted plane condition
is derived, lifting certain assumptions
in previous works. This generalization of the flat plane condition characterizes the total energy surface of a  finite system for all values of electron count $N$ and magnetization $M$. This result is used in combination
with tabulated  spectroscopic data
to show the flat plane-structure
of the oxygen atom,   {among others.  
We find that  derivative discontinuities
with respect to electron count 
sometimes  occur
at non-integer values}.
A diverse set of tilted plane structures is shown to occur in  $d$-orbital subspaces, depending
on  chemical coordination.
General occupancy-based 
  {total-energy} expressions 
 are demonstrated thereby to be 
necessarily dependent 
on the symmetry-imposed degeneracies. 
\end{abstract}

\maketitle


Density Functional Theory (DFT), a workhorse of computational chemistry and condensed matter physics \cite{hohenbergInhomogeneousElectronGas1964a,kohnSelfConsistentEquationsIncluding1965,m.tealeDFTExchangeSharing2022}, owes its success to the development of relatively efficient, reliable and accurate approximations to the exchange-correlation (XC) functional \cite{voskoAccurateSpindependentElectron1980a,perdewGeneralizedGradientApproximation1996a,beckeDensityfunctionalExchangeenergyApproximation1988a,beckeDensityFunctionalThermochemistry1993a,heydHybridFunctionalsBased2003a,leeDevelopmentColleSalvettiCorrelationenergy1988a, perdewGeneralizedGradientApproximation1996b,perdewAccurateSimpleAnalytic1992}. These approximations can be developed by enforcing exact conditions and appropriate norms on a functional of given mathematical form \cite{sunStronglyConstrainedAppropriately2015}. 
Two well-known such exact conditions are the piecewise linearity condition with respect to electron count $N$ \cite{perdewDensityFunctionalTheoryFractional1982a} and the constancy condition with respect to spin-magnetization (henceforth referred to as magnetization for concision) $M$ \cite{yangDegenerateGroundStates2000a,cohenFractionalSpinsStatic2008a} that, 
as we will discuss, is a special case
of a more general linearity condition. 
The former condition states that the total ground state energy of a system with external potential $v(\bf{r})$ and electron count $N=N_0+\omega$ is given by
\begin{equation}
\label{eqn:pwl}
E_v(N_0+\omega)=(1-\omega)E_v(N_0)+\omega E_v(N_0+1),
\end{equation}
where $N_0 \in \mathbb{N}^0$ and $0 \leq \omega \leq 1$~\cite{perdewDensityFunctionalTheoryFractional1982a,yangDegenerateGroundStates2000a,ayersDependenceContinuityEnergy2008}. A DFT calculation for a finite system with a non-integer electron count necessitates fractional occupancy of at least one Kohn Sham (KS) orbital. Assuming this fractional occupancy is limited to one KS orbital, the slope of the linear segment of the $E_v(N)$ curve is given by
\begin{equation}
\label{eqn:janaks_theorem}
\left(\frac{\partial E_v}{\partial N}\right)_v=\epsilon_\textrm{f},
\end{equation}
where $\epsilon_\textrm{f}$ is the fractionally occupied KS eigenvalue \cite{janakProofThatFrac1978,koopmansUeberZuordnungWellenfunktionen1934,perdewCommentSignificanceHighest1997,kronikPiecewiseLinearityFreedom2020}.
A derivative discontinuity must therefore occur at integer values of electron count \cite{perdewPhysicalContentExact1983, shamDensityFunctionalTheoryEnergy1983,sagvoldenDiscontinuityExchangecorrelationPotential2008, yangDerivativeDiscontinuityBandgap2012,  mori-sanchezDerivativeDiscontinuityExchange2014, gouldKohnShamPotentialsExact2014},
  {at least}.
The left-hand partial such 
derivative is given by the highest occupied KS eigenvalue. The right-hand partial derivative is given by the lowest unoccupied KS eigenvalue, $\epsilon_{\rm LUKS}$, with an additional contribution from the derivative discontinuity of the exact XC functional, that is by
\begin{equation}
\label{eqn:janaks_theorem_luks}
\left. \lim_{\delta \to 0^+} \left(\frac{\partial E_v}{\partial N}\right)_v \right|_{N_0+\delta}=\epsilon_{\rm LUKS}+\Delta_\textrm{xc}^N.
\end{equation}
The piecewise linearity condition with respect to electron count \cite{perdewDensityFunctionalTheoryFractional1982a}, as given by Eq.~\ref{eqn:pwl}, follows from the convexity condition  {~\cite{burgessConvexityConditionDensityfunctional2023}}, namely that
\begin{equation}
\label{eqn:convexity_condition}
2E_v[N_0] \leq E_v[N_0-1]+E_v[N_0+1].
\end{equation}
The second of the two aforementioned exact conditions is the constancy condition with respect to magnetization $M$ \cite{yangDegenerateGroundStates2000a,cohenFractionalSpinsStatic2008a}, which states that the total ground state energy of a system with electron count $N_0$ and magnetization $M$ satisfies, for any magnetization $\left| M  \right| \le M_0$,
\begin{align}
\label{eqn:constancy}
&E_v(N_0,M)=E_v(N_0,M_0)=E_v(N_0,-M_0).
\end{align}
Here,  $M_0 \in \mathbb{N}^0$ is the maximum magnetization of the lowest-energy state for a given integer electron count $N_0 \in \mathbb{N}^0$. 
$E_v(N,M)$ denotes the ground state energy of the system with specified electron count $N$ and magnetization $M$, 
noting that these are  independent
parameters in DFT, subject to $ \left| M  \right| \le N$.
Here, and in what follows, it is supposed that no ambient field coupling to spin is present.

The linearity and constancy conditions given by equations \ref{eqn:pwl} and \ref{eqn:constancy}   {have been} combined and generalized \cite{chanFreshLookEnsembles1999,yangDegenerateGroundStates2000a,mori-sanchezDiscontinuousNatureExchangeCorrelation2009a,perdewExactExchangecorrelationPotentials2009,devriendtQuantifyingDelocalizationStatic2021,devriendtUncoveringPhaseTransitions2022,g.janeskoReplacingHybridDensity2021,galEnergySurfaceChemical2010,malekDiscontinuitiesEnergyDerivatives2013,yangCommunicationTwoTypes2016a,cuevas-saavedraSymmetricNonlocalWeighted2012a} to give the flat plane condition
\begin{align}
\label{eqn:flat_plane_condition}
E_v(N,M) ={}& (1-\omega)E_v(N_0,M_0)\nonumber \\ 
{}&+ \omega E_v(N_0+1,M_1)
\end{align}
for $|M| \leq M_0+\omega(M_1-M_0)$, where $N=N_0+\omega$ and $M_1 \in \mathbb{N}^0$ is the maximum magnetization of the lowest-energy state for the $N_0 +1$ electron system.

Many approximate XC functionals   {violate} these exact conditions, and this failure has been directly linked to poor performance in the prediction of band gaps \cite{perdewDensityFunctionalTheory1985a,borlidoLargeScaleBenchmarkExchange2019a,cohenFractionalChargePerspective2008a}, molecular dissociation \cite{ruzsinszkySpuriousFractionalCharge2006a,dutoiSelfinteractionErrorLocal2006a,nafzigerFragmentbasedTreatmentDelocalization2015a,bryentonDelocalizationErrorGreatest}, and electronic transport \cite{toherSelfInteractionErrorsDensityFunctional2005}. A small number of functionals enforce,
fully or to some extent,
the flat plane condition~\cite{burgessMathrmDFTTexttypeFunctional2023,bajajCommunicationRecoveringFlatplane2017,bajajNonempiricalLowcostRecovery2019,janeskoDFTTypeStrong2023,suDescribingStrongCorrelation2018,proynovCorrectingChargeDelocalization2021,kongDensityFunctionalModel2016,johnsonCommunicationDensityFunctional2011,prokopiouOptimalTuningPerspective2022,cuevas-saavedraSymmetricNonlocalWeighted2012a,zhouUnifiedTreatmentDerivative2018,vanaggelenExchangecorrelationEnergyPairing2013}.

  {It has long been recognised in the literature that the constancy condition with respect to magnetisation and the accompanying flat-plane condition apply only to the aforementioned limited interval of magnetization states. In previous
works to generalize this, 
X. Yang et al.~\cite{yangCommunicationTwoTypes2016a} proposed the existence of two types of flat planes, for example, 
and G\'{a}l and Geerlings~\cite{galEnergySurfaceChemical2010} proposed a piecewise linearity condition for magnetization by invoking a zero-temperature grand canonical ensemble.}

  {In this Letter, we  rigorously classify the $E_v[N,M]$ surface for all values of magnetization $M$, as opposed to the limited interval $|M| \leq M_0+\omega(M_1-M_0)$, across all formulations of DFT that satisfy three minimum requirements. These are that the total energy functional is (a) exact for all $v$-representable spin-densities, (b) size-consistent, which is connected to the idea of near-sightedness, and (c)
translationally invariant. We derive a piecewise linearity condition with respect to magnetisation (Thm. 1) and an accompanying, general `tilted plane' condition (Thm. 2). A tilted plane may have a non-zero partial derivative with respect to both $N$ and $M$, while a flat plane (as defined by Eq.~\ref{eqn:flat_plane_condition}) can only have a non-zero partial derivative with respect to $N$. In certain 
cases only, the tilted plane simplifies to the well-known flat plane.}

  {Thms. 1 \& 2 are exact conditions for the XC functional, and so may inform and stimulate the development of improved XC approximations. They highlight the opportunity for functionals that incorporate 
explicit derivative discontinuities with respect to electron count $N$ and magnetization $M$. Thm. 1 opens a new avenue for incorporating spin-polarised atomic reference data as guidance for 
spin-polarized 
functional generalizations~\cite{magyarScalingSpinDensities2002,jacobSpinDensityfunctionalTheory2012,zaharievEnergyComponentsSpindensity2021}.
Furthermore, these two theorems demonstrate that occupancy-based functionals, such as in DFT+U type methods~\cite{petukhovCorrelatedMetalsMathrm2003,parkComputingTotalEnergies2014,burgessMathrmDFTTexttypeFunctional2023,bajajCommunicationRecoveringFlatplane2017,floresClosedLocalorbitalUnified2022,ylvisakerAnisotropyMagnetismTextLSDA2009,janeskoDFTTypeStrong2023} and certain machine-learning based methods~\cite{kirkpatrickPushingFrontiersDensity2021,gedeonMachineLearningDerivative2021,fallettaUnifiedDifferentiableLearning2024}, might beneficially take on different analytical expressions depending on symmetry-imposed degeneracies, as we approach exactitude.}

  {In SM IV, Corollary 1.1 provides the analogue of  the DFT Koopmans' condition for spin, i.e., 
an exact condition on  
the spin-splitting  
of the effective (e.g., Kohn-Sham) 
potential. In spin-polarized systems, 
this splitting underpins the prediction of magnetic excitations and exchange couplings from 
a DFT starting point.}

  {Thm. 2  classifies the $E_v[N,M]$ curve for all values of magnetisation $M$.
Higher-energy magnetisation states are the subject of inherent interest and 
intense research.} For example, the lowest energy triplet state plays a crucial role in phosphorescence \cite{maSupramolecularPurelyOrganic2021,yeConfiningIsolatedChromophores2021}, thermally activated delayed fluorescence \cite{yangRecentAdvancesOrganic2017,amybrydenOrganicThermallyActivated2021} and singlet fission \cite{smithSingletFission2010,buddenSingletExcitonFission2021}. Within   {spin-DFT, such higher energy magnetization states are accessible,} 
as one may compute the lowest energy state of each symmetry 
\cite{gunnarssonExchangeCorrelationAtoms1976}.

{\bf Theorem 1.} {\it The $E_v[N_0,M]$ curve is piecewise linear with respect to magnetization M.}

The structure of the $E_v(N_0,M)$ curve   {(where $N_0\in \mathbb{N}$)} can be elucidated by employing the technique, developed in W. Yang et al.\ \cite{yangDegenerateGroundStates2000a}, 
of constructing a system with external potential $v({\bf r})$ that is composed of $q$ copies of the same finite system with external potential $v_{{\bf R}_l}({\bf r})$, with all copies being infinitely separated in space. 
The duplication of the system, 
at infinite separation,
is a theoretical device used to consider its possible
fractional charge
and spin, without resorting to an ensemble treatment.
The proof proceeds by
analysing the overall  system of duplicates, 
however it is the individual 
finite sub-system (which may be 
strongly multi-reference in character) 
with which we are ultimately concerned. We  have that
\begin{equation}
\label{eqn:external_potential}
v({\bf r})=\sum_{l=1}^q v_{{\bf R}_l}({\bf r}).
\end{equation}
The total electron count is denoted by $qN_0$, where $N_0 \in \mathbb{N}$. It follows from the convexity condition that, at the infinite separation limit, $N_0$ electrons will localize on each of the $q$ sites. For brevity, we will not repeat the electron count constraint
$ \left| M  \right| \le N_0 \in \mathbb{N}_0$, but note that it holds throughout Thm.~1 and Cor.~1.1. The total magnetization of the system is denoted by $qM$, where $q$ and $qM \in \mathbb{Z}$. Typically, $M \notin \mathbb{Z}$ but $M \in \mathbb{Q}$
(which is   {supposed to be} sufficiently dense in $\mathbb{R}$ for present purposes). The ground state of the system will be composed of $p$ sites with a magnetization $M_i$, and $q-p$ sites with a magnetization $M_j$, where
\begin{align}
\label{eqn:total_system_magnetization}
&{}qM = pM_i +(q-p)M_j, \quad
 p,q \in \mathbb{N}^0, \\   &{}p \leq q,  \quad M_i,M_j \in \mathbb{Z}, \quad \mbox{and} \quad M_i \leq M \leq M_j. \nonumber
\end{align}
For fixed values of the external parameters
 $N_0$ and $M$, the total number of sites $q$ is chosen, together with 
  the minimum $M_i$ and maximum $M_j$, such that the total energy per site is minimized while satisfying the constraints of Eq.~\ref{eqn:total_system_magnetization}. Often one finds $M_i=M_j - 2$, but not always. For example, in the nitrogen atom $M_i=-3$ and $M_j=+3$ for $-3<M<+3$. The  choice of $M_i$ and $M_j$ is discussed further in SM I.

  {One of the degenerate ground state wave functions of the system will consist of $p$ sites with magnetisation $M_i$ and $q-p$ sites with magnetisation $M_j$. 
As we describe in detail in 
SM II,
we may take the average of all such ground state wave functions, and analyse the associated spin-resolved site electron density. 
The piecewise linearity condition for magnetisation, i.e.,}
\begin{align}
\label{eqn:magnetization_set_up}
&E_v(N_0,M)=\omega E_v(N_0, M_i)+(1-\omega)E_v(N_0, M_j),  \\ 
&M=\omega M_i+(1-\omega)M_j, {\rm \hspace{0.3cm}} M_i, M_j \in \mathbb{Z} {\rm \hspace{0.3cm}} \& {\rm \hspace{0.3cm}} 0 \leq \omega \leq 1. \nonumber
\end{align}
  {thereby 
follows for all total energy functionals that are (a) exact for all $v$-representable spin-densities, (b) size-consistent, and (c) translationally invariant.}

We note that G\'{a}l and Geerlings \cite{galEnergySurfaceChemical2010} arrived at a similar  piecewise linearity condition for magnetization by invoking a zero-temperature grand canonical ensemble.

\begin{figure}
\includegraphics[width=0.5\textwidth]{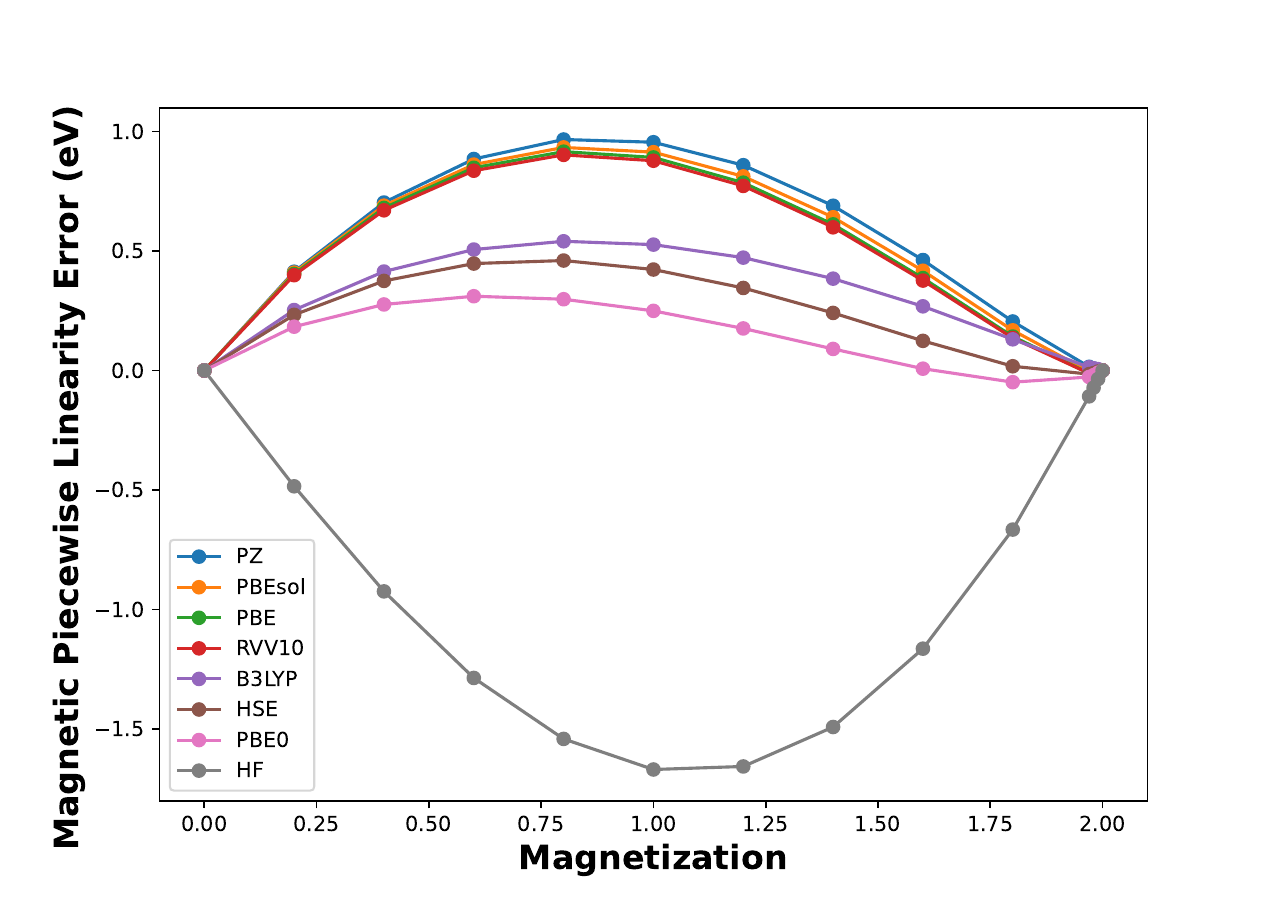}
\caption{The 
magnetic piecewise linearity error 
(MPLE) for the   {neutral Helium atom
exhibited by a variety of local~\cite{perdewSelfinteractionCorrectionDensityfunctional1981}, semi-local~\cite{perdewGeneralizedGradientApproximation1996a,perdewRestoringDensityGradientExpansion2008}, hybrid~\cite{beckeDensityFunctionalThermochemistry1993a,heydHybridFunctionalsBased2003a,leeDevelopmentColleSalvettiCorrelationenergy1988a,perdewRationaleMixingExact1996} and van der Waals corrected~\cite{sabatiniNonlocalVanWaals2013} XC functionals, as well as Hartree-Fock (HF). As defined in Eq.~\ref{mymple}, the MPLE for each approximate functional vanishes at certain integer values of spin-magnetization. MLPE exhibits
a substantial cubic component for all functionals shown. 
It approximately vanishes near full magnetization
(but not elsewhere) for the hybrid functionals, reflecting
the negative sign of MPLE for HF in this test system and hence cancellation
of error  {~\cite{bernardiPhysicalOriginOnequarter2020}}.
 Calculation details~\cite{giannozziQUANTUMESPRESSOModular2009,martynaReciprocalSpaceBased1999,hamannOptimizedNormconservingVanderbilt2013} can be found in SM III.}}
\label{figure:PWL_figure} 
\end{figure}

Approximate total energy functionals $E^{\rm aprx}_v$ typically do not obey the piecewise linearity condition with respect to magnetization. We may refer to the   {intrinsic}
deviation of $E^{\rm aprx}_v$ from the  piecewise linear $E_v(N_0,M)$ curve, 
  {between the (often
inexact) energies at integer
magnetization,}
as magnetic piecewise linearity error (MPLE), 
\begin{align}
\label{mymple}
E^{\rm MPLE}&{}(N_0,M) = E^{\rm aprx}_v(N_0,M) \\ &{} - \left[\omega E^{\rm aprx}_v(N_0,M_i) +(1-\omega)E^{\rm aprx}_v(N_0,M_j)\right]. \nonumber
\end{align}
Here the electron count is a constant integer value $N_0$ and $M$, $M_i$, $M_j$ and $\omega$ are given by equation \ref{eqn:magnetization_set_up}. Fig.~\ref{figure:PWL_figure} shows the $E^{\rm MPLE}(M)$ for  the neutral helium atom.  

MPLE differs from static correlation error (SCE) \cite{cohenFractionalSpinsStatic2008a,burtonVariationsHartreeFock2021}, 
which is defined as the spurious difference in total energy, due to the use of an approximate XC functional, 
between states that should be degenerate.
MPLE instead is an error in the total energy of non-integer magnetization states, irrespective
of whether such a state should be degenerate with one with integer magnetization.
It would not be so, typically, in tilted-plane regions, or in   {a finite} external magnetic field. 
Conversely, not all SCE can be described as MPLE, because SCE may refer to spurious energy differences between 
states of the same magnetization. 
An example of an SCE that is not an MPLE is the energy error that may arise in the 
description of the spherically symmetric boron atom~\cite{cohenFractionalSpinsStatic2008a}.

{\bf Corollary 1.1.} {\it The partial derivative of the $E_v[N_0,M]$ curve with respect to magnetization M is equal to half the   {difference between the spin-dependent frontier KS eigenvalues $\epsilon_\textrm{f}^{\sigma}$}, whenever the partial derivative exists,}
\begin{equation}
\label{eqn:koopmans_for_magnetization_fractional}
\left(\frac{\partial E_v}{\partial M}\right)_N= \frac{1}{2}\left[\epsilon_\textrm{f}^{\upharpoonright}-\epsilon_\textrm{f}^{\downharpoonright} \right].
\end{equation}
A simple proof of this corollary is given in SM IV (see also references~\cite{galEnergySurfaceChemical2010,capelleSpinGapsSpinflip2010,chanFreshLookEnsembles1999,yangDerivativeDiscontinuityBandgap2012,gritsenkoAnalogKoopmansTheorem2002,gritsenkoSpinunrestrictedMolecularKohn2004,galNonuniquenessMagneticFields2009} therein), which avoids the need to invoke total single-particle energies or grand canonical ensembles.

{\bf Theorem 2.} {\it The $E_v[N,M]$ surface obeys the tilted plane condition described by Eq.~\ref{eqn:tilted_plane_condition}.}
\begin{figure*}
\centering
  \includegraphics[width=1.0\textwidth]{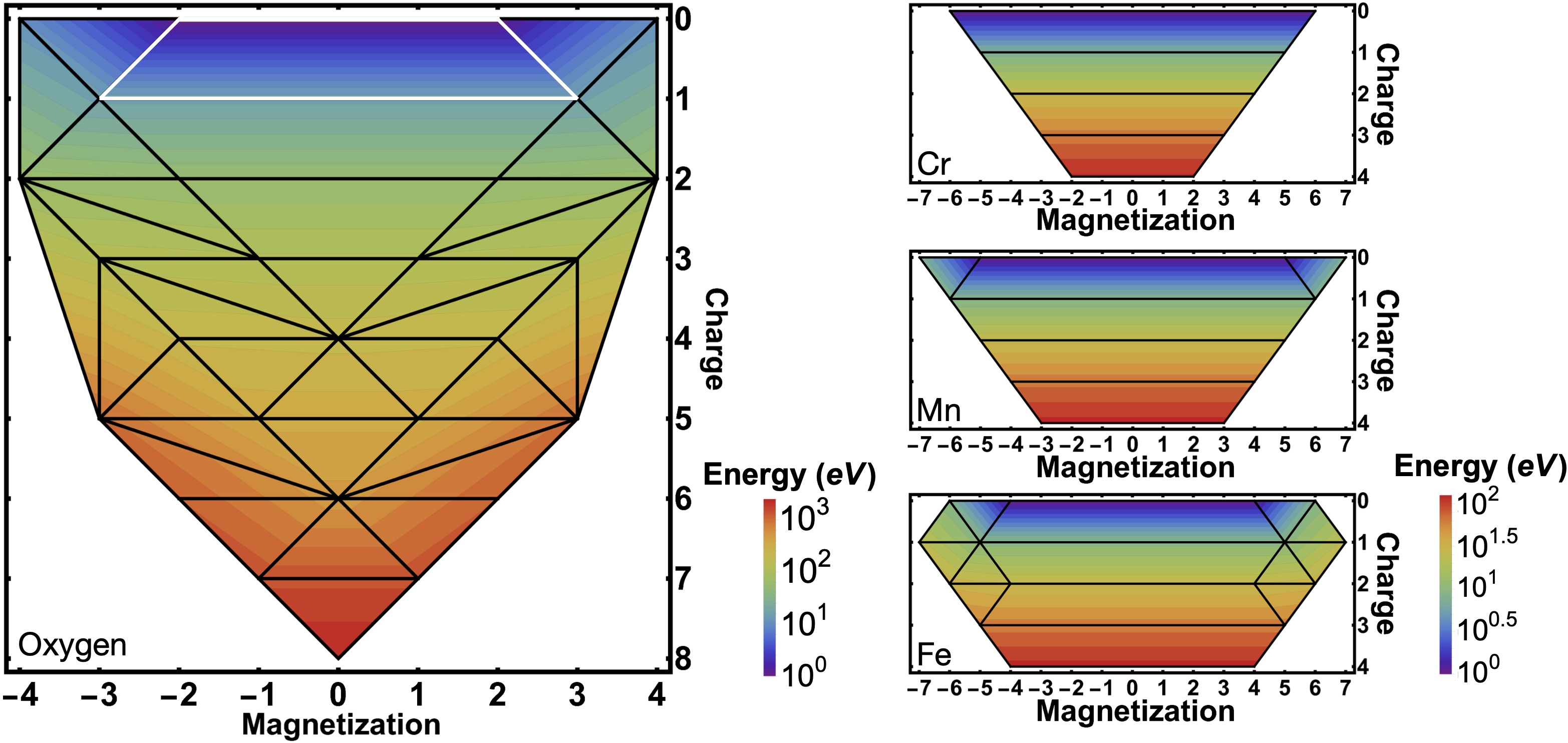}
\caption{The projection of the $E_v[N,M]$ surface of the oxygen, chromium, manganese and iron atoms onto the $N-M$ plane. The $E_v[N,M]$ curves are composed of a series of three- and four-sided planes with vertices at integer values of $N$ and $M$. (The vertical axis is labelled in terms of the total atomic charge as opposed to total electron count.) The total energy varies linearly across each plane.   {Each plane is outlined in black. Energy contour lines are plotted at intervals of $10^{0.1x}$eV.} For each atom, the total energy values are based on available experimental NIST reference data \cite{NIST_ASD} and are given relative to the ground state total energy of the neutral atom in its lowest energy magnetization state,     {which is set to 1~eV for visualization purposes}.
Orbital and nuclear moments are not included.
  {The $E_v[N,M]$ surface of the oxygen atom for charged states between $+5$ and $+6$ is composed of a series of five triangular shaped planes, four of which can neither be classed as a type 1 or a type 2 flat plane of X. Yang et al.~\cite{yangCommunicationTwoTypes2016a}.} 
Note that planes at the edges are omitted due to absence of experimental reference data. If any additional experimental reference data points not currently included are found to be low enough in energy, they could also affect the shape of the 
  {outlying, 
high-magnetization} $E_v[N,M]$ surfaces shown here.}

\label{figure:oxygen}
\end{figure*}

Analysis of the $E_v[N_0,M]$ curve may be extended to include states with not only fractional magnetization $M$ but also fractional electron count $N$. Again, one may construct an external potential given by Eq.~\ref{eqn:external_potential}, in this case typically both $N$ and $M \notin \mathbb{Z}$, but the total system electron count and magnetization, $qN$ and $qM \in \mathbb{Z}$. The ground state of this system will be composed of $qc_i$ sites with electron count $N_i$ and magnetization $M_i$, where $i$ ranges from 1 to $V_{\rm count}$ with:
\begin{equation}
 \sum_{i=1}^{V_{\rm count}} c_i=1,  {\rm \hspace{0.4cm}} 0  \leq c_i \leq 1,   {\rm \hspace{0.4cm}} N_i,{\rm \hspace{1.0mm}} M_i,{\rm \hspace{1.0mm}} qc_i \in \mathbb{Z}.
\end{equation}
The values of $c_i$, $N_i$ and $M_i$ are constrained so that
\begin{equation}
N=\sum_{i=1}^{V_{\rm count}} c_i N_i \quad \mbox{and} \quad M=\sum_{i=1}^{V_{\rm count}} c_iM_i.
\end{equation}
Following an analogous derivation to that outlined in Thm.~1   {(inc. SM I and II),} one finds that, 
for $N,M \notin \mathbb{Z}$, 
\begin{equation}
\label{eqn:tilted_plane_condition}
E_v[N,M]=\sum_{i=1}^{V_{\rm count}} c_iE_v[N_i,M_i].
\end{equation}

Vertices in the energy landscape will occur at the specified integer values of electron count $N_i$ and magnetization $M_i$. $V_{\rm count}$ is the number of vertices associated with a particular plane, typically  equal to 3 or 4, however higher numbers of vertices are possible in very rare circumstances, as discussed in SM V.   {To simplify   discussion,} we restrict $V_{\rm count}=4$, 
however, the 
method used to generate Fig.~\ref{figure:oxygen} includes no restriction on $V_{\rm count}$. 

In cases where the electron counts of the four states 
  {at the vertices} 
satisfy $N_1=N_2=N_3-1=N_4-1$ and the magnetizations  $M_1=-M_2$,  $M_3=-M_4$, with $E_v[N_i,M_i]=\min_M\{E_v[N_i,M]\}$, Eq.~\ref{eqn:tilted_plane_condition} simplifies to the flat-plane condition as outlined in Eq.~\ref{eqn:flat_plane_condition}. X. Yang et al.~\cite{yangCommunicationTwoTypes2016a} and Cuevas-Saavedra et al.~\cite{cuevas-saavedraSymmetricNonlocalWeighted2012a} report the existence of two types of flat plane structures. 
  {These types  are 
special cases 
(non-exhaustive ones; 
see Fig.~\ref{figure:oxygen}
for an exception)} 
of the more general condition 
outlined by Eq.~\ref{eqn:tilted_plane_condition}, specifically when $c_4=0$.   {Generally,}  restricting $c_4$ to be zero prohibits the correct flat plane structure of systems, e.g., the oxygen atom for $7 \leq N\leq 8$ and $|M| \leq 10-N$,   {as highlighted in white in Fig.~\ref{figure:oxygen}. The} correct expression for   {this energy surface is detailed in SM VI, along with further analysis of Fig.~\ref{figure:oxygen}.}

G\'{a}l and Geerlings \cite{galEnergySurfaceChemical2010} reported the existence of a tilted plane energy surface but their energy expressions also have the $c_4=0$ restriction, meaning that $(N_i,M_i)$ values in their energy expression will not always represent vertices in the energy landscape. The same is true for Chan's $E_v[N,M]$ energy expression \cite{chanFreshLookEnsembles1999}. To the best of our knowledge, a lifting of the $c_4 = 0$ restriction of the generalized flat plane condition has only been discussed to date in the unpublished Ref.~\cite{malekDiscontinuitiesEnergyDerivatives2013}. If we assume that $(N_i,M_i)$ values represent vertices in the energy landscape, the $c_4=0$ restriction allows for triangular shaped planes but neglects planes of other shapes, such as isosceles trapezoids, which   {occur in the oxygen atom (Fig.~\ref{figure:oxygen}).}

  {A consequence 
worthy of future investigation 
is the appearance of derivative 
discontinuities both at non-integer
values of total electron count $N$,
and of spin-electron count $\frac{1}{2}(N\pm M)$, 
as seen, e.g., in the oxygen atom  
between 
charge states 5+ and 6+ with $1\le M\le2$ (Fig.~\ref{figure:oxygen}).}
Knowledge of the occurrence of a tilted plane energy surface has already been applied to correct the PBE total energy of dissociated triplet H$_5^+$ in Ref.~\cite{burgessMathrmDFTTexttypeFunctional2023}.
An analysis of the tilted plane shape of the $E_v[N,M]$ surface for the He atom  is shown in SM III and the $E_v[N,M]$ surface for the 
multi-reference, quintuple bonded VCr, VMo and CrMo hetero-dimers~\cite{ruiperezElectronicStructureBonding2011} are presented in SM VII. Sample degeneracy-dependent tilted plane structures for an   {isolated} $d$-orbital subspace are shown in SM VIII.

In conclusion, the piecewise linearity condition with respect to magnetization and the tilted plane condition were derived from first principles using the infinite-separation-limit technique, and the magnetic analogue of the DFT Koopmans’ theorem was derived from the chain rule.   {These exact conditions have been derived for all formulations of DFT that are exact for all $v$-representable spin-densities, size-consistent, and translationally invariant. It is important to note that these exact conditions apply only to the lowest $E_v[N,M]$ energy surface. By performing a constrained search over states of a specified spin multiplicity~\cite{gunnarssonExchangeCorrelationAtoms1976}, it is possible to access other quantum states within DFT that are not located on this energy surface, for example, the lowest energy singlet state of a system with a triplet ground state.} 

  {We have found that many standard density functional approximations violate these exact conditions.} These three exact quantum mechanical conditions   {may} aid in the development of post-DFT methods and functional approximations, 
machine learning and alchemical approaches in
both condensed matter physics 
and quantum chemistry, 
and error-correction techniques 
involving total energies in quantum science. 
We find that in order to approach
the exact limit, energy functionals of occupancies
must necessarily take different forms depending on 
symmetry-imposed degeneracies.  

\begin{acknowledgments}
The research conducted in this publication was funded by the Irish Research Council under grant number GOIPG/2020/1454. EL gratefully acknowledges financial support from the Swiss National Science Foundation (grant number 213082). This research was supported by the NCCR MARVEL, a National Centre of Competence in Research, funded by the Swiss National Science Foundation (grant number 205602).
\end{acknowledgments}

\bibliography{main}

\clearpage

\renewcommand{\theequation}{S\arabic{equation}}
\renewcommand{\thefigure}{S\arabic{figure}}
\renewcommand{\thetable}{S\arabic{table}}
\setcounter{equation}{0}
\setcounter{figure}{0}

\section{SM-I. The correct choice of site magnetizations}
\label{sec:BLOR-derivation}
 {In the derivation of the piecewise linearity condition for magnetization, 
care must be taken in determining the site magnetizations $M_i$ and $M_j$.
Here we have given values of the external parameters $N_0$ and $M$, where $qN_0$ and $qM$ are, 
respectively, the total electron count and the total magnetization of the system composed of $q$ infinitely separated sites.} The minimum $M_i$ and maximum $M_j$ values which minimize the total energy per site, while satisfying the constraint of Eq.8,  are determined by  {the following}  three conditions.
\begin{enumerate}
\item
The site magnetizations $M_i$ and $M_j$ must satisfy
\begin{equation}
\label{eqn:condition_inequality}
M_i\leq M \leq M_j \quad M_i,M_j \in \mathbb{Z}.
\end{equation}
    \item
For any values of site magnetization $M_h, M_k \in \mathbb{Z}$ where $M_i \notin \{M_h, M_k\}$, the ground-state energy of the system with magnetization $M_i$ must satisfy
\begin{align}
\label{eqn:condition2_on_mi}
&{\rm \hspace{1cm}}E_{v_{{\bf R}_l}}(M_i) < c_hE_{v_{{\bf R}_l}}(M_h)+c_kE_{v_{{\bf R}_l}}(M_k), \\
&{\rm \hspace{1cm}}M_i =c_hM_h+c_kM_k,{\rm \hspace{0.3cm}} c_h, c_k \geq 0 {\rm \hspace{0.3cm}}\&{\rm \hspace{0.3cm}} c_h+c_k=1. \nonumber 
\end{align}
The equivalent condition for $M_j$ must also be true.
    \item 
For any integer value of site magnetization $M^{\prime}$ in the range $M_i \leq M^{\prime} \leq M_j$, the ground-state energy must satisfy
\begin{align}
\label{eqn:condition1_on_mi}
&{\rm \hspace{1cm}}E_{v_{{\bf R}_l}}(M^{\prime}) \geq c_iE_{v_{{\bf R}_l}}(M_i)+c_jE_{v_{{\bf R}_l}}(M_j), \\
&{\rm \hspace{1cm}}M^{\prime}=c_iM_i+c_jM_j,{\rm \hspace{0.3cm}} c_i, c_j \geq 0 {\rm \hspace{0.3cm}}\&{\rm \hspace{0.3cm}} c_i+c_j=1. \nonumber 
\end{align}

Thus, the same values of $M_i$ and $M_j$ will satisfy these criteria for all values of the external parameter $M$ that satisfy Eq.~\ref{eqn:condition_inequality}.
\end{enumerate}

 {\section{SM-II. The Piecewise Linearity Condition for Magnetization}}
 {Given a system of $q$ infinitely separated sites with total electron count $qN_0$ and magnetization $qM$, where the ground state of the system is composed of $p$ sites with a magnetization $M_i$, and $q - p$ sites with a magnetization
$M_j$, subject to the constraints 
of Eq. 8 in the main text, we explicate in full the proof of Eq. 9 in the main text, which specifies the piecewise linearity condition for magnetization. The correct choice in values of $M_i$ and $M_j$ have already been discussed in SM-I.}

 {The system has a set of degenerate ground state wave functions. At least one of these degenerate wave functions will be the antisymmetrized product of ground state wave functions at each site [14,16]. This follows because in the infinite separation limit, each site cannot interact with its neighbours and, thus, minimizing the energy of the total system is equivalent to separately minimizing the energy associated with each site subject to the constraints 
of Eq. 8.
One such ground state will have the first $p$ sites with magnetization $M_i$ and wave function $\Phi_{M_i}$, and the remaining $q-p$ sites with magnetization $M_j$ and wave function $\Phi_{M_j}$. The ground state wave function of the total system is} 
 {\begin{align}
\label{eqn:groud_state_wavefunction_psi1}
\Psi_1= \hat{A} \left(
\Phi_{M_i} \right. & ({\bf{R}}_1)   \cdots\Phi_{M_i}
({\bf{R}}_p)  \\ \nonumber & \left.  \Phi_{M_j}
({\bf{R}}_{p+1})\cdots\Phi_{M_j}({\bf{R}}_q)
\right).
\end{align}
It is important to note that the ground state wave function of Eq.~\ref{eqn:groud_state_wavefunction_psi1} holds regardless of whether each finite 
sub-system can be well-described by a single-determinant wave function or not.}

 {Exchanging the magnetizations $M_i$ and $M_j$ at any two sites results in a degenerate ground state wave function. Therefore, the average of all such wave functions $\Psi_{{\rm avg}}$ is also a ground state wave function of the system. The ground state spin $\sigma$ density of $\Psi_{{\rm avg}}$ is given by
\begin{equation}
\label{eqn:spin-density-expression}
\rho^{\sigma}({\bf r})=\sum_{l=1}^q \frac{p}{q} {\rm \hspace{0.75mm}} \rho_{l}^{\sigma }({\bf r};M_i)+\frac{q-p}{q} {\rm \hspace{1mm}} \rho_{l}^{\sigma}({\bf r};M_j),
\end{equation}
where $\rho_{l}^{\sigma }({\bf r};M_i)$ is the spin $\sigma$ spin density of site $l$ with a magnetization of $M_i$. In this case, each of the $p$ (and separately  $q-p$)  sites have identical spin-resolved densities. We empahsize that Eq.~\ref{eqn:spin-density-expression} is one of the degenerate ground state spin $\sigma$ densities of the system with total magnetization $qM$, where $q$ is the number of sites and $M$ is limited to the interval $M_i \leq M \leq M_j$.}

 {To deduce the piecewise linearity condition for magnetization, we make three reasonable assertions about the nature of the total energy functional, namely that it is (a) exact for all $v$-representable spin-densities, (b) size-consistent, and (c) translationally invariant. 
We expect that the exact  functional satisfies these in all DFT formulations where the ground state is generally accessible.}

 {From (a), the total energy function should be exact for the spin resolved density of Eq.~\ref{eqn:spin-density-expression}, so that
\begin{equation}
\label{eqn:exact-energy}
E_v[\rho^{\sigma }
]=pE_{v_{{\bf R}_l}}(M_i)+(q-p)E_{v_{{\bf R}_l}}(M_j).
\end{equation}
From (b), the total energy functional should be size-consistent, whereupon
\begin{equation}
\label{eqn:size-consistent-energy}
E_v(\rho^{\sigma }
)=\sum_{l=1}^qE_{v_{{\bf R}_l}}\left(\frac{p}{q} {\rm \hspace{0.75mm}} \rho_{l}^{\sigma }(
M_i)+\frac{q-p}{q} {\rm \hspace{1mm}} \rho_{l}^{\sigma }(
M_j)\right).
\end{equation}
Eq.~\ref{eqn:size-consistent-energy} can be simplified by application of (c), translational invariance, following which
\begin{equation}
\label{eqn:translational-invariance-energy}
E_v[\rho^{\sigma }
]=qE_{v_{{\bf R}_l}}\left[\frac{p}{q} {\rm \hspace{0.75mm}} \rho_{l}^{\sigma }(
M_i)+\frac{q-p}{q} {\rm \hspace{1mm}} \rho_{l}^{\sigma }(
M_j)\right].
\end{equation}
From Eqs.~\ref{eqn:exact-energy} and \ref{eqn:translational-invariance-energy}, the total energy of the isolated site $l$ with magnetization $M_l$ is given by
\begin{align}
\label{eqn:energy_of_site_l}
E_{v_{{\bf R}_l}}(N_0,M_l)=&\frac{p}{q}E_{v_{{\bf R}_l}}(N_0, M_i)\nonumber \\ &+\frac{q-p}{q}E_{v_{{\bf R}_l}}(N_0, M_j),
\end{align}
where the site magnetization is given by
\begin{equation}
\label{eqn:site_magnetization}
M_l=\frac{p}{q}M_i+\frac{q-p}{q}M_j=M.
\end{equation}
An equivalent argument will hold for any value of $M$ in the range $M_i \leq M \leq M_j$. Making the change of variable $\omega =p/q$ and relabelling the site potential ${v_{{\bf R}_l}}({\bf r})$ simply as $v({\bf r})$, we may succinctly state that
\begin{align}
\label{eqn:magnetization_set_up}
&E_v(N_0,M)=\omega E_v(N_0, M_i) 
\\ \nonumber &\quad\quad\quad\quad
\quad\;\;\;
+ (1-\omega)E_v(N_0, M_j),  \\ 
&M=\omega M_i+(1-\omega)M_j, {\rm \hspace{0.3cm}} M_i, M_j \in \mathbb{Z} {\rm \hspace{0.3cm}} \& {\rm \hspace{0.3cm}} 0 \leq \omega \leq 1. \nonumber
\end{align}
Therefore, a spin density functional $E_v[\rho^{\upharpoonright }({\bf r}),\rho^{\downharpoonright }({\bf r})]$ 
satisfying the three aforementioned conditions obeys a piecewise linearity condition with respect to magnetization, as opposed to simply a constancy condition.}

\vspace{0.6cm}

 {\section{SM-III. Helium atom at Fractional Magnetization}}

\begin{figure}
\centering
\includegraphics[scale=0.5]{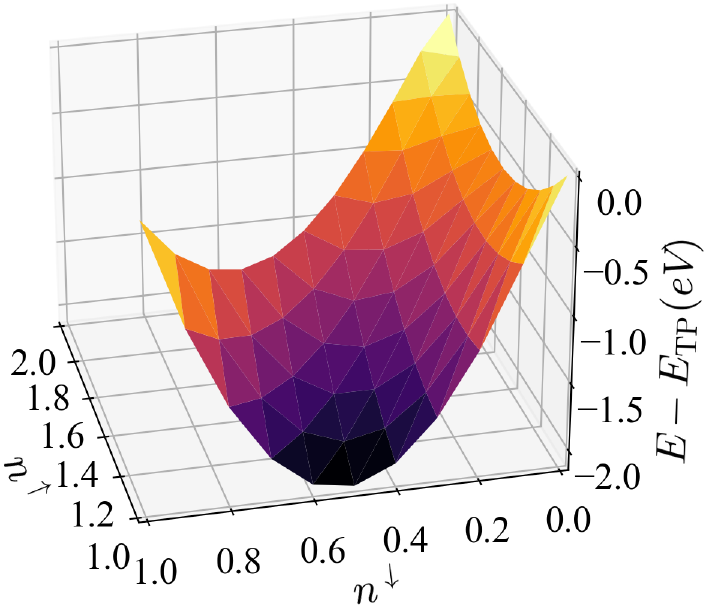}
\caption{A plot of the deviation of the total energy using the PBE XC functional from the tilted plane condition for the Helium atomic system with spin up and spin down electron counts in the range $1\leq n^{\upharpoonright} \leq 2$ and $0\leq n^{\downharpoonright} \leq 1$, assuming the PBE total energy is exact at integer values of $n^{\upharpoonright}$ and $n^{\downharpoonright}$.}
\label{figure:tilted_plane}
\end{figure}

 {It is instructive to apply our analysis on the helium energy surface at fractional values of $N$ and $M$. }The $E_v[N,M]$ surface for the helium atom with electron count in the range $1 \leq N\leq 2$ and magnetization $M \geq 2-N$ is given by
\begin{align}
\label{eqn:he_tilted_plane}
E_v[N,M] ={\mbox{ }} & (2-N)E_v[1,1]+\frac{N-M}{2}E_v[2,0] \nonumber \\
& + \frac{M+N-2}{2}E_v[2,2].
\end{align}
This plane will have a non-zero partial derivative with respect to magnetization and thus has a `tilted plane' shape as opposed to a simple `flat plane' shape.
Approximate XC functionals typically do not obey the tilted plane condition. For example, Fig.~\ref{figure:tilted_plane} displays the energy of the helium atom when using the PBE XC functional, 
with respect to the exact tilted plane outlined in Eq.~\ref{eqn:he_tilted_plane}.

 {All  DFT calculations
on the helium atom, as presented also in Fig. 1 of  the main text, were performed using the pw.x code of the
Quantum Espresso package [79]. Kintetic-energy 
cutoffs of 75 Ry and 300 Ry were set for the Kohn Sham wavefunctions and charge density, respectively. Martyna-Tuckerman periodic boundary corrections~[80] were applied, along with a Gaussian 
occupancy smearing of 0.02 Ry. Davidson diagonalization was employed with Broyden charge density mixing, a mixing factor of 0.7, and a convergence threshold of $5\times 10^{-9}$ Ry. The helium atom was positioned in the centre of a $20\times 20\times20$~\AA$^3$ cell, with $\Gamma$-point only reciprocal space sampling. The Optimized Norm-Conserving Vanderbilt helium pseudopotential [81] from the SG15 collection was used, which is available at www.quantum-simulation.org. In the case of Hartree-Fock, B3LYP, HSE and PBE0, the fully magnetized $(M=2)$ helium calculations would not converge. In this case, the total energy at $M=2$ was evaluated by linear extrapolation of the total energies from the $M=1.97,$ $1.98$ \& $1.99$ calculations. }

\clearpage

\begin{figure}[h]
\centering
\includegraphics[scale=0.5]{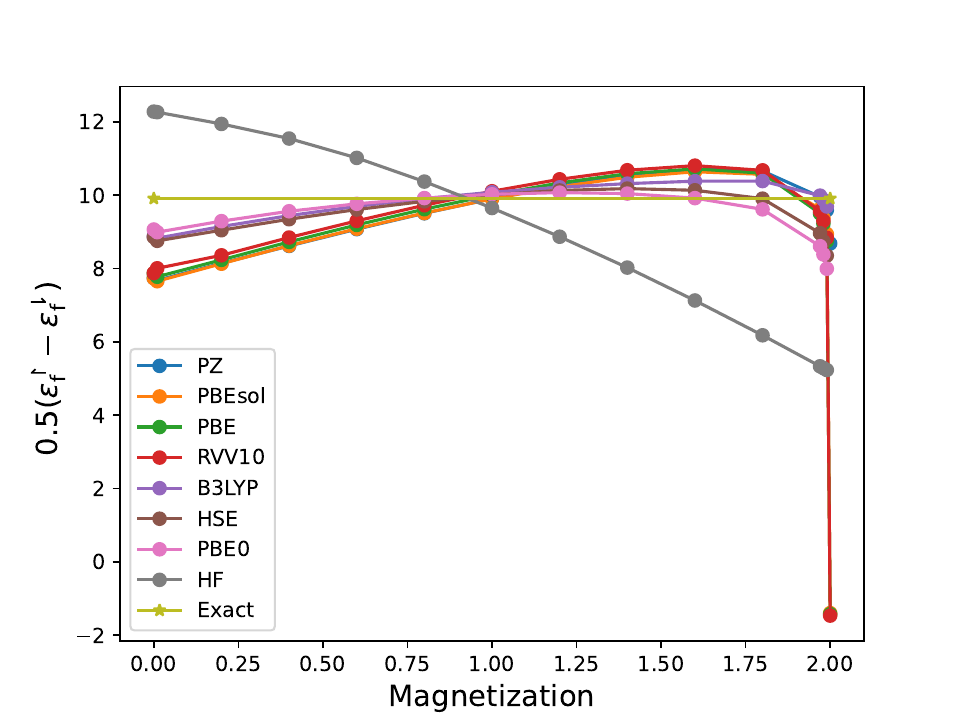}
\caption{Half the difference of the fractionally occupied KS eigenvalues for the He atom, plotted as a function of magnetization $M$ {, for a variety of local [74], semi-local [5,75], hybrid  [7–9, 76] and van der Waals corrected XC functionals [77], as well as Hartree-Fock (HF)}.  The exact  { constant line, based on  experimental NIST reference data [88],} is also displayed. The  {lack of a derivative discontinuity within certain XC approximations}  can result in very large deviations from the magnetic
analogue of Koopmans’ theorem at certain integer values of $M$, as is the case shown here  {for the PBEsol, PBE and RVV10 functionals at $M= 2.0$. As discussed in SM-III, the hybrid functionals as well as Hartree-Fock were evaluated at $M=1.97$, $1.98$ \& $1.99$ but not at $M= 2.0$.}}
\label{figure:koopmans_figure}
\end{figure}

\section{SM-IV. The magnetic analogue of the DFT Koopmans’ theorem}
{\bf Corollary 1.1.} {\it The partial derivative of the $E_v[N_0,M]$ curve with respect to magnetization 
M is equal to half the  {difference between the spin-dependent frontier KS eigenvalues $\epsilon_\textrm{f}^{\sigma}$}, whenever the partial derivative exists.}

The magnetic analogue of the DFT Koopmans’ theorem can be derived by simple application of the chain rule in conjunction with the well studied (spin resolved) DFT Koopmans’ theorem. This dispenses with the need to invoke total single particle energies or grand canonical ensembles used in previous proofs [34, 84] of this theorem. The partial derivative of the total energy with respect to magnetization may be expressed in terms of the spin resolved electron counts
\begin{equation}
\label{eqn:partial_deriv_decomposition}
\left(\frac{\partial E_v}{\partial M}\right)_{ N}=\sum_{\sigma}
\left(\frac{\partial E_v}{\partial N^{\sigma}}\right)_{N^{\bar{\sigma}}}
\left(\frac{\partial N^{\sigma}}{\partial M}\right)_N ,
\end{equation}whenever the necessary partial 
derivatives exist. The right and left  partial derivatives of $E_v$ with respect to $N^{\sigma}$ are given, respectively,  by
\begin{align}
\label{eqn:deriv_with_respect_to_nsigma}
\lim_{\delta \to 0^+} \left.\left(\frac{\partial E_v}{\partial N^{\sigma}}\right)\right|_{N^{\sigma}_0 + \delta} &= \epsilon_{\rm LUKS}^{\sigma} +\Delta_\textrm{xc}^{N^{\sigma}} {\mbox{ and}} \nonumber \\
\lim_{\delta \to 0^-} \left.\left(\frac{\partial E_v}{\partial N^{\sigma}}\right)\right|_{N^{\sigma}_0 + \delta} &= \epsilon_{\rm HOKS}^{\sigma},
\end{align} where 
$\epsilon_{\rm HOKS}^{\sigma}$ is the highest occupied spin-$\sigma$ KS eigenvalue and $\epsilon_{\rm LUKS}^{\sigma}$ is the lowest unoccupied spin-$\sigma$ KS eigenvalue. $\Delta_\textrm{xc}^{N^{\sigma}}$ is the explicit derivative discontinuity of the exact XC functional through its explicit dependence on $N^{\sigma}$. Eq.~\ref{eqn:deriv_with_respect_to_nsigma} is the spin resolved analogue of Eqs.~[2] and [3]. For further details,  see Refs.~[24, 28, 85-87].

The right and left partial derivatives of the total energy with respect to magnetization of a system with no fractional KS orbital occupations is thus given by
\begin{align}
\label{eqn:koopmans_for_magnetization}
\lim_{M\to M_0^+} \left(\frac{\partial E_v}{\partial M}\right)_N &= \frac{1}{2}\left[\epsilon_{\rm LUKS}^{\upharpoonright}-\epsilon_{\rm HOKS}^{\downharpoonright} + \Delta_\textrm{xc}^{N^{\upharpoonright}}\right] \quad {\mbox{and}} \nonumber \\
\lim_{M\to M_0^-} \left(\frac{\partial E_v}{\partial M}\right)_N &= \frac{1}{2}\left[\epsilon_{\rm HOKS}^{\upharpoonright}-\epsilon_{\rm LUKS}^{\downharpoonright} - \Delta_\textrm{xc}^{N^{\downharpoonright}}\right],
\end{align}since 
the partial derivative of $N^{\sigma}$ with respect to $M$ is equal to one half. For systems with one fractionally occupied spin up 
frontier KS orbital of eigenvalue $\epsilon_\textrm{f}^{\upharpoonright}$, 
and one fractionally occupied spin down one
of eigenvalue $\epsilon_\textrm{f}^{\downharpoonright}$, the expression given by Eq.~\ref{eqn:koopmans_for_magnetization} simplifies to
\begin{equation}
\label{eqn:koopmans_for_magnetization_fractional}
\left(\frac{\partial E_v}{\partial M}\right)_N= \frac{1}{2}\left[\epsilon_\textrm{f}^{\upharpoonright}-\epsilon_\textrm{f}^{\downharpoonright} \right].
\end{equation} Typical 
XC functionals will break this exact condition. In Fig. \ref{figure:koopmans_figure} the slope of the $E_v[N_0,M]$ curve for the Helium atom, as evaluated from the PBE KS eigenvalues, is plotted as a function of magnetization. Use of the exact XC functional would result in a perfect step function.

\section{SM-V. Cases of high vertex count}
In special cases, the vertex count of a given plane can be higher than four. For example, let us assume that the maximum magnetization of the lowest energy state of the $N-1$, $N$ and $N+1$ electron systems are given by $M$, $M+1$, and $M$,  respectively, where $M \geq 1$. It is assumed that the convexity condition of Eq. 4 is satisfied for the $N-1$, $N$ and $N+1$ electron systems. In the special case where
\begin{align}
\label{eqn:fortuitous_slopes}
&{}\frac{E_v[N,M+1]-E_v[N-1,M]}{N-(N-1)} \\
\nonumber &{}\quad=
\frac{E_v[N+1,M]-E_v[N,M+1]}{(N+1)-N},
\end{align}
the plane in question will have an hexagonal shape (stretched in the magnetization direction whenever $M>1$). Such a plane would require the summation in Eq. 14 to be extended over six vertices 
\begin{equation}
E_v[N,M]=\sum_{i=1}^6 c_i  E_v[N_i,M_i].
\end{equation}

Higher numbers of vertices will occur when there is a fortuitous equality of certain energy derivatives. Eq.~\ref{eqn:fortuitous_slopes} specifies the energy derivative equality for this particular example. It is worth noting that Eq.~\ref{eqn:fortuitous_slopes} simplifies to
\begin{equation}
2E_v[N,M+1]=E_v[N-1,M]+E_v[N+1,M],
\end{equation}
however this simplification will not occur in all cases of high vertex count.

\begin{figure*}[t]
\centering
  \includegraphics[width=1.0\textwidth]{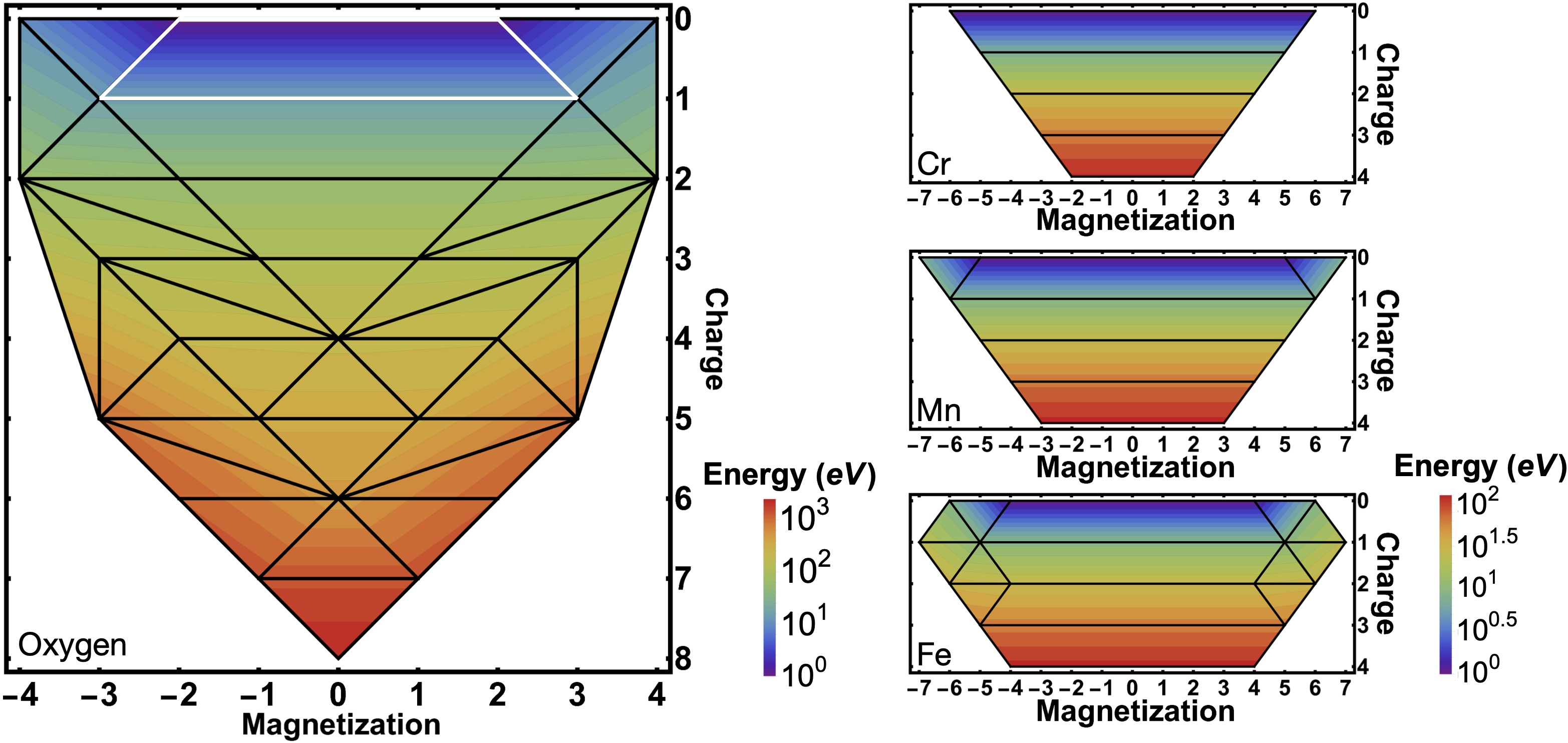}
\caption{ {The projection of the $E_v[N,M]$ surface of the oxygen, chromium, manganese and iron atoms onto the $N-M$ plane. The $E_v[N,M]$ curves are composed of a series of three- and four-sided planes with vertices at integer values of $N$ and $M$. (The vertical axis is labelled in terms of the total atomic charge as opposed to total electron count.) The total energy varies linearly across each plane. Each plane is outlined in black. Energy contour lines are plotted at intervals of $10^{0.1x}$eV. For each atom, the total energy values are based on available experimental National Institute of Standards and Technology reference data [88] and are given relative to the ground state total energy of the neutral atom in its lowest energy magnetization 
state, which is set to 1~eV for visualization purposes.
Orbital and nuclear moments are not included.
The $E_v[N,M]$ surface of the oxygen atom for charged states between $+5$ and $+6$ is composed of a series of five triangular shaped planes, four of which can neither be classed as a type 1 or a type 2 flat plane of X. Yang et al. [36].
 Within this region, the energy surface has been incorrectly produced by G\'{a}l and Geerlings in Fig. 3 of their work [34], highlighting the
challenge of  mapping these planes.
Note that planes at the edges are omitted due to absence of experimental reference data. If any additional experimental reference data points not currently included are found to be low enough in energy, they could also affect the shape of the 
outlying, 
high-magnetization $E_v[N,M]$ surfaces shown here.
The flat plane described by Eq.~\ref{eqn:oxygen_flat_plane} is highlighted in white.} }
\label{figure:oxygen2}
\end{figure*}

 {
\section{SM-VI. Further Analysis of Figure 2}}

 {A number of other interesting observations can be made from Fig. 2 in the main text, for simplicity we have included the same figure here, labelled Fig.~\ref{figure:oxygen2}. If we look, for example, at the 5+ oxygen anion, we can read from the graph that the energy  is minimized at
$M=1$ (the spin-doublet configuration 1s$^2$2s$^1$), with a piecewise constant segment for $\left| M \right| \le 1 $,
while the next lowest magnetization state is  $M=3$ (the spin-quartet configuration 1s$^1$2s$^1$($^3$S)2p$^1$), with piecewise linear (not piecewise constant) segments connecting these for $1 \le \left| M \right| \le 3 $. }

 {It is also worth noting that the neutral chromium atom violates the Madelung rule, with a ground state electron configuration of [Ar]3d$^5$4s$^1$. This can been seen in the central uppermost plane of the $E_v[N,M]$ surface for Cr, which extends from $-6$ to $+6$ at zero charge as opposed to $-4$ to $+4$, which would be the case if the configuration followed the Madelung rule ([Ar]3d$^4$4s$^2$).}

 {The correct energy expression for the oxygen atom, without the $c_4=0$ restriction, may
be written as }
\begin{widetext}
 {\begin{align}
\label{eqn:oxygen_flat_plane}
E_v[N,M]=&c_1E_v[8,2]+c_2E_v[8,-2]+c_3E_v[7,3]+c_4E_v[7,-3]=\frac{(N-7)(10-N+M)}{2(10-N)}E_v[8,2]+ \\ & \frac{(N-7)(10-N-M)}{2(10-N)}E_v[8,-2]+ \frac{(8-N)(10-N+M)}{2(10-N)}E_v[7,3]+\frac{(8-N)(10-N-M)}{2(10-N)}E_v[7,-3], \nonumber
\end{align}}
\end{widetext}
 {for all $7 \leq N \leq 8$ and $|M| \leq 10 - N$. The flat plane described by Eq.~\ref{eqn:oxygen_flat_plane} is highlighted in white in Fig.~\ref{figure:oxygen2}.} 


 {Despite $E_v[8,2]=E_v[8,-2]$ and $E_v[7,3]=E_v[7,-3]$ for the oxygen atom, reduction  of Eq.~\ref{eqn:oxygen_flat_plane} to the sum of two terms with coefficients $c_1^{\prime}$ and $c_2^{\prime}$ would  wrongly give 
\begin{equation}
M \neq \sum_{i=1}^2 c_i^{\prime} M_i.
\end{equation}
The black lines within the $E_v[N,M]$ surfaces of Fig. 2 (Fig.~\ref{figure:oxygen2}), correspond to values of $N$ \& $M$ where derivative discontinuities in the total energy occur. As shown in the $E_v[N,M]$ surface of the oxygen atom, these derivative discontinuities do not solely occur at integer values of $N$ \& $M$, or at integer values of the spin up and spin down electron counts ($N^{\upharpoonright}$ and $N^{\downharpoonright}$). }

 {For example, there is a line segment
stretching from the point $(0,6)$ to the point $(3,5)$.  
This line represents a `seam' on which, in this case, both of the partial derivatives change.
Thus, there is a derivative discontinuity in the total energy for all values of total charge $ C = 8 - N $ (8 being the atomic number of oxygen) and total magnetization $M$ on this line segment.
The equation of the line segment is $C=-\frac{1} {3}M+6$, and so derivative discontinuies occur, e.g.,  at $M=1.5$ and $C=5.5$ as well as at the integer magnetization $M=2$ and $C=5 \frac{1}{3}$.
This discontinuity (for varying $N$
at constant $M$)
has a magnitude of approximately 10.2~eV.}

 {This example also provides 
that a derivative discontinuity may be 
encountered while
one  spin occupancy alone is  varied,
at possibly non-integer values.
Recall that the spin-indexed occupancies are 
given by 
$N^\upharpoonright = \left( N + M \right) / 2$ and 
$ N^\downharpoonright = \left( N - M \right) / 2$.
We may consider traversing from point $(2,6)$
(which corresponds to $N^\upharpoonright = 2$ and
$N^\downharpoonright = 0$) to 
point $(1,5)$
(which corresponds to $N^\upharpoonright = 2$ and
$N^\downharpoonright = 1$). 
A derivative discontinuity is encountered
at half-integer filling 
$N^\downharpoonright = 0.5$,
which is at the aforementioned point $(1.5,5.5)$, 
half-way along the line segment.
It is worth noting that there appears to be
no possible absent NIST spectroscopic data 
in the region in question that might correspond
to higher-energy spin states,
the existance of which might
otherwise change the fracturing of the $E_v[N,M]$
landscape and affect this prediction.}

\clearpage

\section{SM-VII.  $\mathbf{E_v[N,M]}$  surface of sample transition metal hetero-dimers}

\begin{figure}
\centering
\includegraphics[scale=0.08]{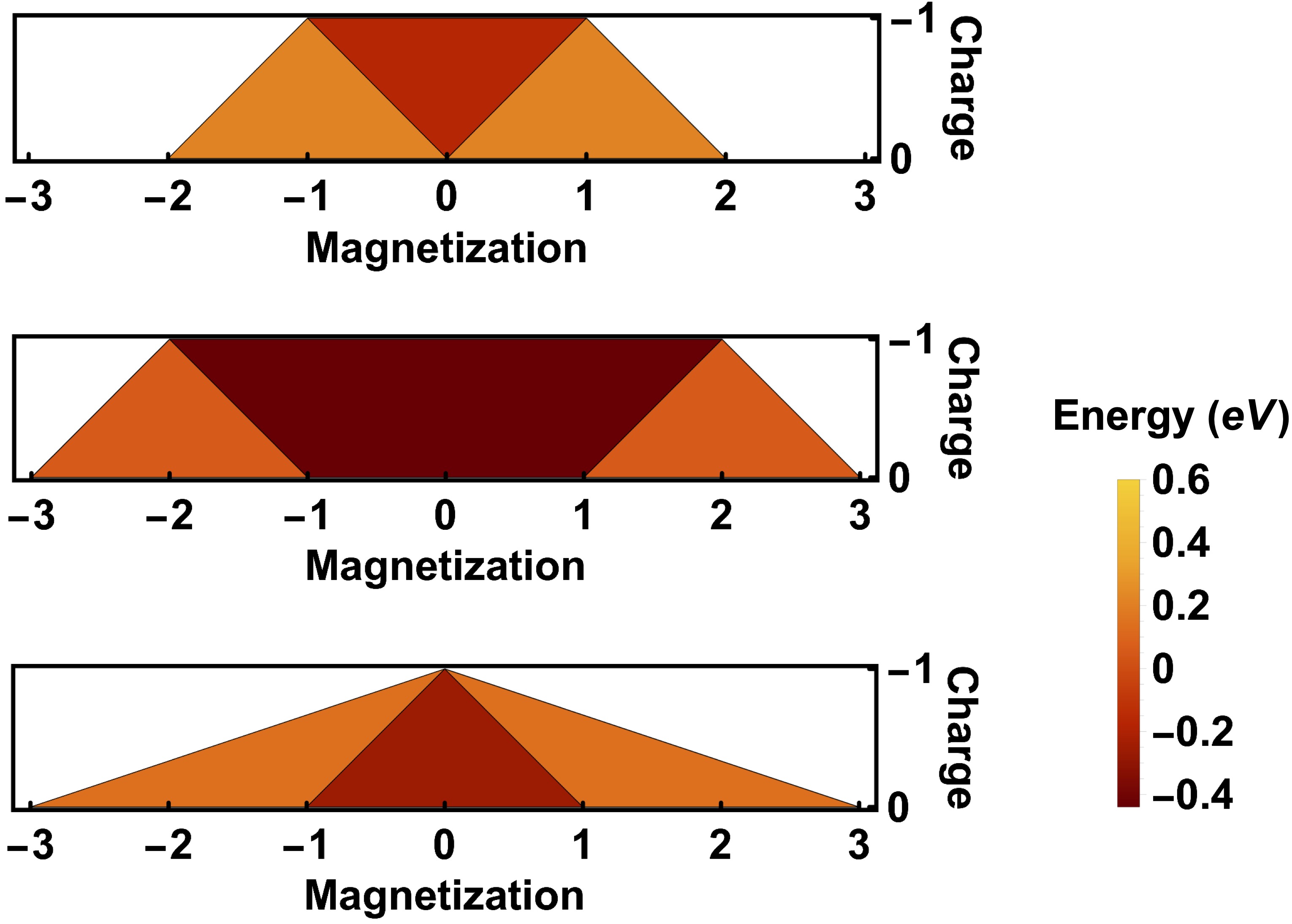}
\caption{The projection of the $E_v[N,M]$ surface onto the $N-M$ plane for the CrMo (A), VCr (B) and VMo (C) hetero-dimers based on CASSCF/CASPT2 reference calculations for vertical charge and spin excitations [89]. The y-axis is labelled in terms of total charge as opposed to total electron count.}
\label{figure:heterodimers}
\end{figure}

The projection of the $E_v[N,M]$ surface onto the $N-M$ plane for the strongly correlated CrMo, VCr and VMo transition metal hetero-dimers are presented in Fig~\ref{figure:heterodimers}. In particular, please note that the $E_v[N,M]$ surfaces are composed of a series of three and four sided planes with vertices at integer values of $N$ and $M$, as is the case with the oxygen, chromium, manganese and iron atoms. The total energy varies linearly across each plane but, for ease of visualization, each plane is colored based on the average of the energies at the vertices. The total energy values for the strongly correlated CrMo, VCr and VMo transition metal hetero-dimers are based on CASSCF/CASPT2 reference calculations on vertical excitations from  Ref. [89]. 
This study classified the ground state
of each dimer, in both the neutral and
anionic state, as having a pronounced 
multi-reference character, as well as an effective bond order
indicative of a quintuple bond.

For each  system, the total energy values are relative to the ground state total energy of the neutral dimer in its lowest energy magnetization state. Note that only a subset of the full $E_v[N,M]$ surface is presented for these systems, this is due to the limited data available. If any additional reference data points not currently included are found to be low enough in energy, they could also affect the shape of the $E_v[N,M]$ surfaces shown here.

 {\section{SM-VIII.  Model crystal/ligand field-dependent tilted plane structures for a $\bf d$-orbital subspace}}

\begin{figure*}
  \includegraphics[width=\textwidth]{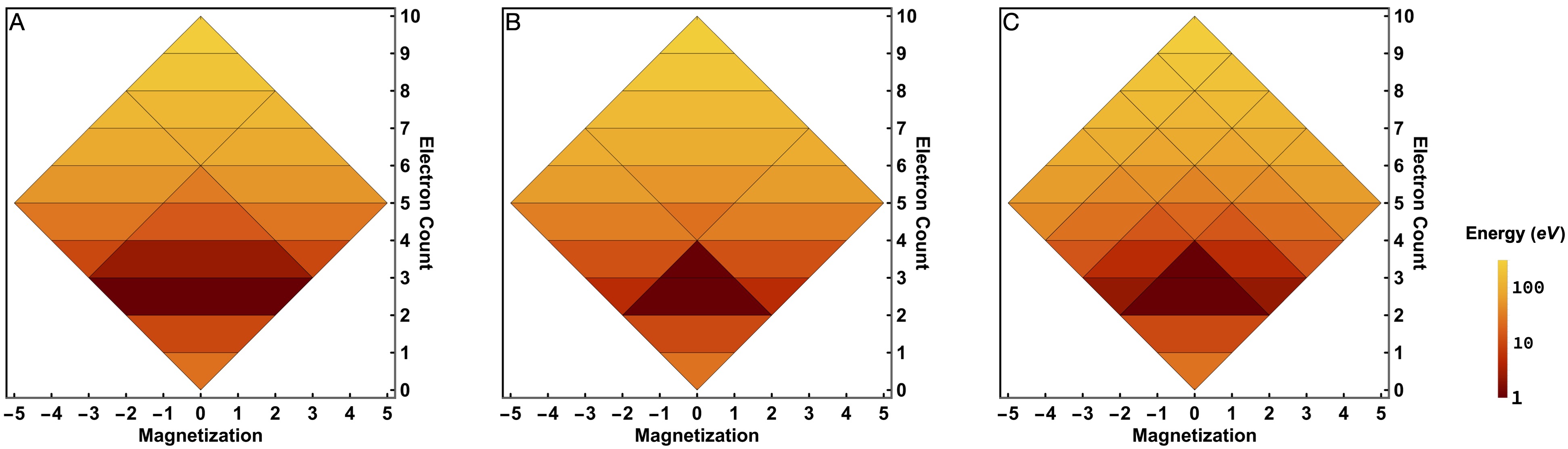}
\caption{Sample tilted plane structures for a $d$-orbital subspace in the subspace-bath linear interaction limit, under octahedral (A), tetrahedral (B) and square-planar (C) crystal field splittings. In the model energy (see Eq.~\ref{eqn:model_energy}), 
parameters were chosen to show a sample
of the 
the rich coordination-dependence of 
flat plane shapes.
The interaction parameters $U$ and $V$ were set at 11 eV and 10 eV, respectively, and the orbital energy levels were chosen so that $\sum_i\epsilon_i=-80$ eV. The set of orbital energy levels $\{-24,-24,-24,-4,-4\}$, $\{-25,-25,-10,-10,-10\}$ and $\{-25,-25,-18,-10,-2 \}$, in eV, were used for the octahedral, tetrahedral and square-planar crystal fields, respectively. The total energy varies linearly across
each plane but, for ease of visualization, each plane is colored based on the average of the energies at the vertices. In each example, the energies are reported relative to the plane with lowest average energy, whose energy was set to 1eV to allow a logarithmic color scale.}
\label{figure:tilted_planes-d-orbitals}
\end{figure*}

The energies of localized electronic states within a solid material also obey the tilted plane condition in the limit where the subspace-bath interaction energy varies linearly with spin resolved occupancy. This of course includes the case where the subspace-bath interaction becomes negligible. Fig.~\ref{figure:tilted_planes-d-orbitals} displays sample tilted plane structures for a $d$-orbital subspace in the subspace-bath linear interaction limit under octahedral, tetrahedral, and square-planar crystal field splittings. The total energy of the electronic states with integer spin-resolved orbital occupancies were approximated using the model 
\begin{equation}
\label{eqn:model_energy}
E^{\rm mod}=\sum_i \epsilon_i n_i+U \sum_i n_i^{\upharpoonright}n_i^{\downharpoonright}+\frac{V}{2}\sum_{i \neq j}n_i n_j,
\end{equation}
where  {$U$ and $V$ are the intra-orbital and inter-orbital interaction parameters, respectively. The values $\epsilon_i$ and ${n}_i^{\sigma}$
are the  energy level and
 spin-$\sigma$ electron count, respectively, 
of orbital $i$.} 
For each crystal field splitting, and 
hence degeneracy pattern, a wide variety of possible tilted pane structures exist. 
 {Total-energy approximations  
 defined for non-integer spin-occupancies
arise in a wide variety of 
DFT-related methods, e.g., 
in DFT+$U$ type XC-corrections and in machine learning proxies 
for total-energy components.
These and perhaps next-generation 
XC approximations, as we approach exactitude, 
could beneficially  
(recalling Fig.~\ref{figure:tilted_planes-d-orbitals})
(a) change expressions discretely between different 
tilted-plane segments, and (b) 
change in the manner of that change, 
depending on symmetry-imposed degeneracy 
patterns.}

\end{document}